\documentclass[10pt]{article}
\usepackage[left=2cm,top=2cm,right=2cm,nohead]{geometry}
\usepackage{amsmath,amssymb,amsthm}
\usepackage{mathtools}

\usepackage{graphicx}
\usepackage{color}
\usepackage{authblk}

\newcommand{\cnot}[0]{{\sc{cnot}}}
\newcommand{\cz}[0]{{\sc{cz}}}

\title{A simple method for sampling random Clifford operators}
\author{Ewout van den Berg}
\affil{IBM Quantum, IBM T.J.~Watson Research Center\\ Yorktown Heights, NY, USA}

\begin{document}

\maketitle

\begin{abstract}
  We describe a simple algorithm for sampling $n$-qubit Clifford
  operators uniformly at random. The algorithm outputs the Clifford
  operators in the form of quantum circuits with at most $5n + 2n^2$
  elementary gates and a maximum depth of $\mathcal{O}(n\log n)$ on
  fully connected topologies. The circuit can be output in a streaming
  fashion as the algorithm proceeds, and different parts of the
  circuit can be generated in parallel. The algorithm has an
  $\mathcal{O}(n^2)$ time complexity, which matches the current
  state of the art. The main advantage of the proposed algorithm,
  however, lies in its simplicity and elementary derivation.
\end{abstract}

\section{Introduction}

The $n$-qubit Clifford group, $\mathcal{C}_n$, consists of all unitary
operators for which the signed $n$-qubit Pauli group, $\mathcal{P}_n$,
is closed under conjugation. Operators from the Clifford group can be
implemented as quantum circuits consisting of only Hadamard, phase,
and controlled-not (\cnot) gates. Conversely, any circuit made up of
only these and derived gates, such as single-qubit Pauli gates and
\cz\ gates, so-called Clifford circuits, implements an element of the
Clifford group. An important property of Clifford circuits is that
they can be efficiently simulated when the initial state is a
computational basis state~\cite{AAR2004Ga}. In addition to simulation
and various other applications~\cite{BRA2020Ma}, Clifford operators
play an important role in the characterization of noise channels using
techniques such as randomized
benchmarking~\cite{EME2005AZa,KNI2008LRBa,MAG2011GEa}.  Some of these
applications, randomized benchmarking included, require that we can
efficiently sample elements from $\mathcal{C}_n$ uniformly at random.

The first efficient algorithm for sampling from the Clifford group was
given by Koenig and Smolin~\cite{KOE2014Sa}. They start with the
observation that the cardinality of $\mathcal{C}_n$ is finite:
\begin{equation}\label{Eq:CardinalityC}
\vert\mathcal{C}_n\vert = 2^{n^2+2n}\prod_{j=1}^n (4^j-1).
\end{equation}
It then follows that randomly sampling of the group is equivalent to
sampling an integer index between $0$ and $\vert\mathcal{C}_n\vert-1$
and providing a one-to-one mapping between indices and the elements of
$\mathcal{C}_n$. As a next step, they observe that
$\mathcal{C}_n / \mathcal{P}_n \cong Sp(2n,\mathbb{F}_2)$, the
symplectic group on $\mathbb{F}_2^{2n}$. Given that the mapping from
integers to the Pauli group is trivial, it remains to find a mapping
to elements of the symplectic group. Efficient algorithms for this and
the inverse mapping based on transvections are given
in~\cite{KOE2014Sa}, resulting in a sampling algorithm with time
complexity $\mathcal{O}(n^3)$. In recent work, Bravyi and
Maslov~\cite{BRA2020Ma} study the structure of the Clifford group and
provide a canonical representation for elements in the group based on
the Bruhat decomposition~\cite{MAS2018Ra}. Leveraging this canonical
form, they obtain an $\mathcal{O}(n^2)$ algorithm for sampling
elements from the Clifford group uniformly at random. They also
describe how variants of the canonical form can be used to implement
arbitrary Clifford unitaries with a circuit depth of at most $9n$ on a
linear nearest-neighbor architecture.

In this work we provide a simple algorithm for uniform sampling from
the Clifford group. The algorithm is based on the tableau
representation of Paulis and, similar to~\cite{KOE2014Sa}, takes
advantage of the hierarchical structure of the Clifford group.  The
algorithm has an $\mathcal{O}(n^2)$ runtime and, unlike other
algorithms, directly generates circuits with depth at most
$\mathcal{O}(n\log n )$ on a fully-connected topology. The circuits can
be generated in a streaming fashion with $\mathcal{O}(n)$ latency and
segments of the circuit can be generated fully in parallel. Perhaps
the main advantage of the algorithm, however, lies in its simplicity and
elementary derivation.
The algorithm leverages the tableau representation introduced
in~\cite{AAR2004Ga}. We describe this convenient representation in
detail in Section~\ref{Sec:Tableau}. The motivation behind the
proposed algorithm is given in Section~\ref{Sec:Motivation}, followed
by a derivation of the algorithm in Section~\ref{Sec:Algorithm}. We
conclude with a brief discussion in Section~\ref{Sec:Discussion}.

\section{Paulis and tableau representation}\label{Sec:Tableau}

\begin{figure}
\centering
\begin{tabular}{cccc}
\raisebox{12.5pt}{\includegraphics[width=0.22\textwidth]{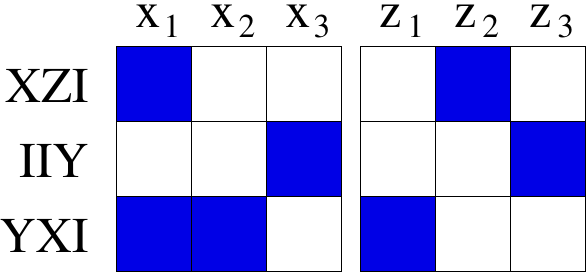}}\
  &
\includegraphics[width=0.23\textwidth]{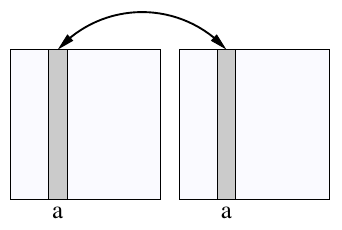}&
\includegraphics[width=0.23\textwidth]{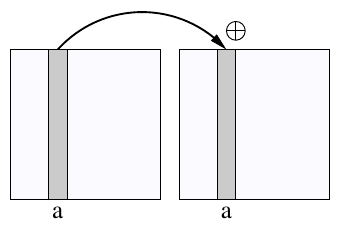}&
\includegraphics[width=0.23\textwidth]{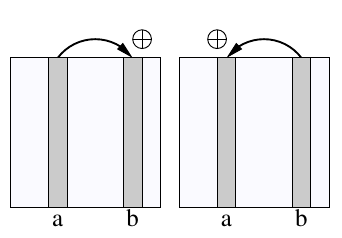}\\
\ \ \ Tableau  & H(a) &
S(a) & CX(a,b)
\end{tabular}
\caption{Tableau representation of Paulis with X and Z blocks (left),
  along with the effect of conjugating with single-qubit operators H
  and S (center) and the two-qubit operator CX. In this figure we
  omit the sign vector associated with the Paulis, as well as the
  updates to this sign vector resulting from the conjugation
  operations.}\label{Fig:Tableau}
\end{figure}

Pauli operators are formed as tensor products of the $2\times 2$
identity matrix $I$, and the three Pauli matrices
\[
X = \left[\begin{array}{cc}0&1\\1&0\end{array}\right],\quad
Y = \left[\begin{array}{cc}0&-i\\i&0\end{array}\right],\quad
Z = \left[\begin{array}{cc}1&0\\0&-1\end{array}\right].
\]
Elements from the $n$-qubit Pauli group are generated using $n$
components and can be written as $2^n\times 2^n$ operators
$P = \otimes_i P_i$. Multiplying two Pauli operators reduces to
multiplying the corresponding components: given a second Pauli
operator $Q = \otimes_i Q_i$, we have $PQ = \otimes _i (P_iQ_i)$.  An
important property of the Pauli group is that any two elements either
commute ($PQ=QP$) or anticommute ($PQ = -QP$). Each Pauli matrix
commutes with itself and the identity matrix, and anticommutes with
the other two Pauli matrices. It follows from the multiplication rule
given above that elements $P$ and $Q$ anticommute if and only if an
odd number of components $P_j$ and $Q_j$ anticommute. Denote by $X_j$
the tensor product $\otimes_i P_i$ with $P_j =X$ and $P_i=I$ for all
$ i\neq j$, and define $Y_j$ and $Z_j$ is a similar manner. We can
then conveniently represent Paulis in terms of binary vectors
$x, z \in \mathbb{F}_2^n$:
\begin{equation}\label{Eq:PauliXZ}
P(x,z) = \prod_j i^{x_jz_j} X_j^{x_j}Z_j^{z_j}.
\end{equation}
The factors $i^{x_jz_j}$ are added to correct for the phase
resulting from the multiplication $X_jZ_j = -iY_j$. We can represent a
set of $k$ Paulis as a $k\times 2n$ binary matrix, or tableau, where
each row contains the $x$ and $z$ coefficients of a single
Pauli. The leftmost plot in Figure~\ref{Fig:Tableau} illustrates a
tableau representing the Pauli operators $XZI = X\otimes Z\otimes I$,
$IIY$, and $YXI$ with the $x$ and $z$ coefficients grouped together
into $X$ and $Z$ blocks. The order of the columns in the tableau can
be changed to suit our needs and we sometimes use tableaus in which
the $x$ and $z$ coefficients alternate. Tableaus can additionally be
augmented with a column of sign bits $s$ that indicate a phase
$(-1)^s$. Although all tableaus in the paper will have such a sign
column, we sometimes omit them from illustrations to keep the
exposition clean.

As mentioned in the introduction, the Clifford group is generated by
the single-qubit Hadamard (H) and phase (S) gates, and the two-qubit
controlled-not gate (CX). Any quantum circuit consisting of only these
and derived gates (including all Pauli operators and the controlled-Z
gate) implements a Clifford operator
$\mathcal{C}(\rho) = C\rho C^{\dag}$. (Note that such a circuit
implementation is by far unique: any given Clifford operator has
infinitely many circuit representations.) The power of the tableau
notation lies in the ease with which it represents the mapping of
Pauli operators under Clifford operators. For instance, as illustrated
in Figure~\ref{Fig:Tableau}, applying the Hadamard gate on qubit $a$
results in the exchange of the corresponding columns in the
tableau. Conjugation with the phase gate results in addition of $x_a$
to $z_a$, modulo two, for each of the Paulis in the tableau, and
finally, application of CX on qubits $a$ and $b$ adds $x_a$ to $x_b$
and $z_b$ to $z_a$.  Each of these operations has an associated update
to the sign vector, as detailed in~\cite{AAR2004Ga}.
Since Clifford operators can be written as products of unitary
operators, they themselves must be unitary. Given Clifford
operator $\mathcal{C}$ and any two Pauli operators $P$ and $Q$ we
therefore have
\[
\mathcal{C}(PQ)
= C(PQ)C^{\dag}
= CP(C^{\dag}C)QC
= \mathcal{C}(P)\mathcal{C}(Q).
\]
This means that if we know the Pauli operators resulting from
conjugation of $P$ and $Q$, we know the resulting operator for Pauli
$PQ$. We know from~\eqref{Eq:PauliXZ} that any Pauli can be written as
a product of basis terms $X_j$ and $Z_j$, so once we know
$\mathcal{C}(X_j)$ and $\mathcal{C}(Z_j)$ for all $j$, we can
determine the action of $\mathcal{C}$ on all other Paulis. In order to
fully describe a Clifford operator it thus suffices to prescribe the
mapping of the $2n$ basis terms. These mappings have a number of
restrictions. First, the identity always maps to itself:
$CIC^{\dag} = ICC^{\dag} = I$. Second, the mapping is a bijection; no
two Paulis map to the same Pauli. Third, commutation relations between
elements remain invariant under conjugation. In other words, if Paulis
$P$ and $Q$ commute then so do $P'=\mathcal{C}(P)$ and
$Q'=\mathcal{C}(Q)$:
\begin{align*}
0 &= PQ - QP = C(PQ-QP)C^{\dag}\\
&= CP(C^{\dag}C)QC^{\dag} - 
CQ(C^{\dag}C)PC^{\dag}  = P'Q' - Q'P'.
\end{align*}
Note that the conjugation of a Pauli with a Clifford
operator can results in a signed Pauli. For instance, 
\hbox{$HYH^{\dag} = -Y$}.

\begin{figure}
\centering
\begin{tabular}{cccc}
\includegraphics[width=0.2\textwidth]{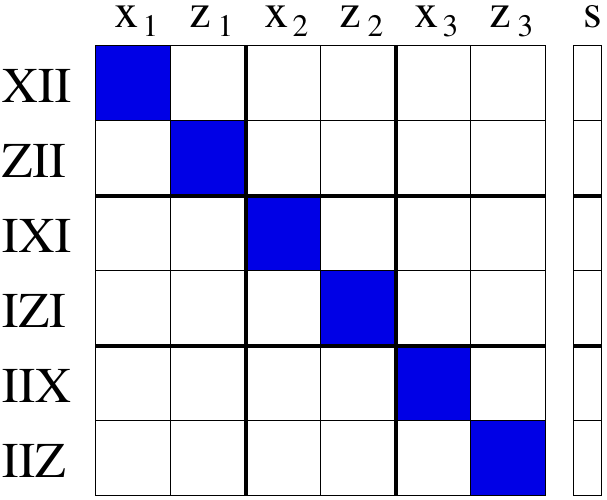}&
\includegraphics[width=0.2\textwidth]{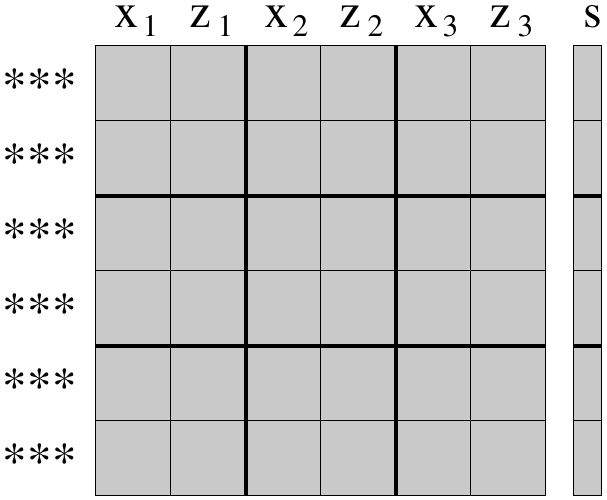}&
\includegraphics[width=0.2\textwidth]{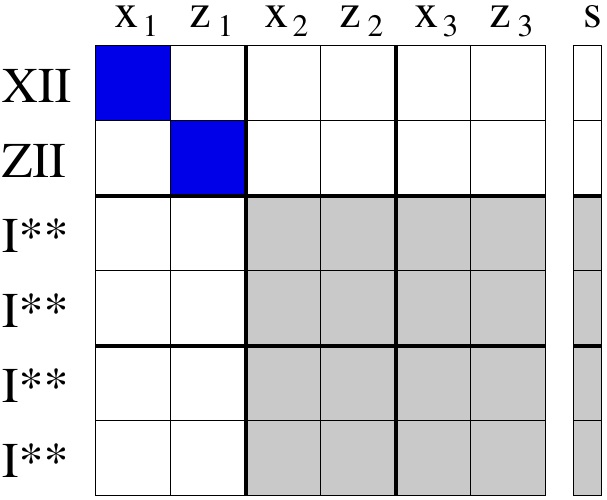}&
\includegraphics[width=0.2\textwidth]{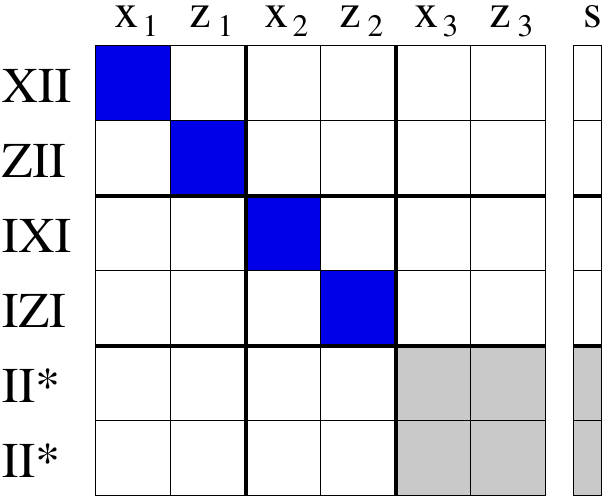}
\\
({\bf{a}}) & ({\bf{b}}) & ({\bf{c}}) & ({\bf{d}})
\end{tabular}
\caption{Tableau representation of (a) the Pauli basis, and (b) the
  Paulis that result by applying Clifford operator $\mathcal{C}$ to
  the Pauli basis tableau. Gray entries indicate possibly
  non-zero elements and the asterisk symbols each indicate any one of the $I$,
  $X$, $Y$, or $Z$ components. This is the initial tableau for
  subsequent sweeping operations. The state of the tableau (c) after
  sweeping the first pair of rows using a suitable Clifford operator
  $\mathcal{C}_1$, and (d) after sweeping the second pair of rows
  using $\mathcal{C}_2$.}\label{Fig:TableauA}
\end{figure}

\section{Motivation}\label{Sec:Motivation}

Consider an initial tableau that contains the interleaved set of basis
Paulis $X_j$ and $Z_j$, as shown in Figure~\ref{Fig:TableauA}(a). By
applying Clifford operator $\mathcal{C}$ we obtain a new tableau that
contains Paulis $\mathcal{C}(X_j)$ and $\mathcal{C}(Z_j)$, along with
their sign. Since the new tableau specifies the output of the map from
all basis Paulis, it completely represents the Clifford
operator. Suppose now that we are given a tableau corresponding to a
random Clifford operator $\mathcal{C}$, as illustrated in
Figure~\ref{Fig:TableauA}(b), where gray entries represents possibly
non-zero elements, and where an asterisk (*) denotes any one of the
$I$, $X$, $Y$, or $Z$ components. We can convert the tableau back to a
Clifford operator as follows. First, using an appropriate combination
of operations from Figure~\ref{Fig:Tableau} on qubits 1,2, and 3, we
can normalize the first two rows and obtain the tableau as shown in
Figure~\ref{Fig:TableauA}(c). The successive operators applied to the
tableau correspond to Clifford operators, and we denote their product
by $\mathcal{C}_1$. Repeating a similar sweeping procedure on qubits 2
and 3, we generate Clifford operator $\mathcal{C}_2$ and obtain the
tableau given in Figure~\ref{Fig:TableauA}(c). Finally, operations on
qubit 3 with operator $\mathcal{C}_3$ results in the basis tableau
shown in Figure~\ref{Fig:TableauA}(a). Given that the tableau in
Figure~\ref{Fig:TableauA}(b) was determined by applying the random
operator $\mathcal{C}$ on the basis tableau, it follows that the
reverse operator $\mathcal{C}^{\dag}$ is given by
$\mathcal{C}_3\mathcal{C}_2\mathcal{C}_1$. Taking the adjoint of this
operator therefore allows us to obtain the original operator
$\mathcal{C}$ (the circuit representation of the operator could
differ).

\section{Proposed algorithm}\label{Sec:Algorithm}

\begin{figure}[t]
\centering
\begin{tabular}{cccc}
\includegraphics[width=0.21\textwidth]{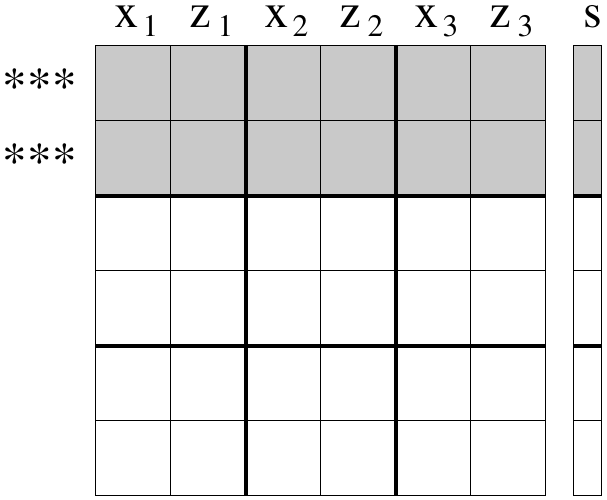}&
\includegraphics[width=0.21\textwidth]{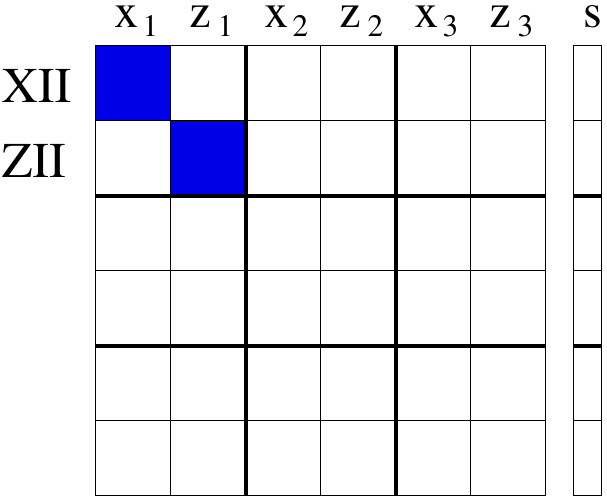}&
\includegraphics[width=0.21\textwidth]{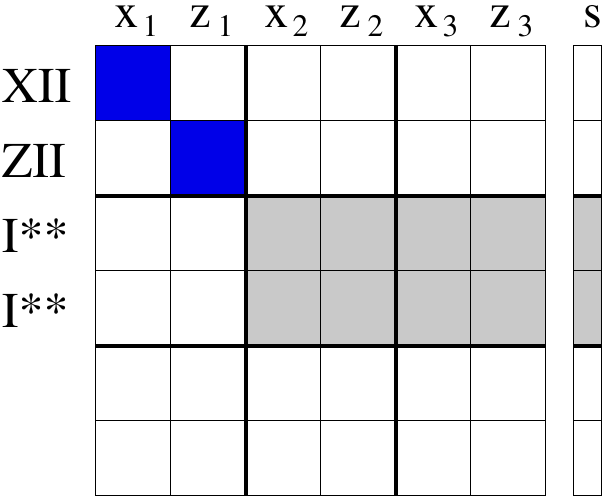}&
\includegraphics[width=0.21\textwidth]{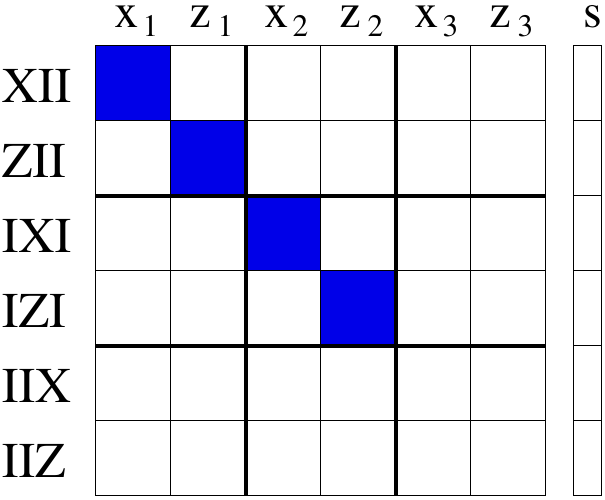}\\
\ \ ({\bf{a}}) & \ \ ({\bf{b}}) &
\ \ ({\bf{c}}) & \ \ ({\bf{d}})
\end{tabular}
\caption{(a) The tableau with the first two rows initialized with
  random anticommuting Paulis; (b) the state after sweeping the
  first two rows of the tableau to basis states $X_1$ and $Z_1$; (c)
  start of the second iteration with random Paulis on the second pair
  of rows; and (d) state after sweeping these rows to $X_2$ and $Z_2$.}\label{Fig:TableauB}
\end{figure}

The proposed algorithm is similar to the approach described in the
previous section but operates on two rows at a time rather than on the
full tableau. We initialize an empty tableau with $n$ rows and
populate the first two rows with randomly selected anticommuting Pauli
operators and random signs, as shown in
Figure~\ref{Fig:TableauB}(a). As before, we then manipulate these rows
such that the first Pauli becomes $X_1$ and the second becomes $Z_1$,
as illustrated in Figure~\ref{Fig:TableauB}(b).
If we had applied to same manipulations on a fully populated tableau
shown in Figure~\ref{Fig:TableauA}(b), we would now have the tableau
shown in Figure~\ref{Fig:TableauA}(c). However, instead of sampling a
full tableau and updating it with every sweep, we simply randomly
sample what the next two rows would have been after sweeping. The
crucial point here is that it does not matter whether we first sample
the rows randomly at the beginning and the sweep, or first sweep and
then randomly sample what they would have been after the
transformation. To be consistent with the result of the sweeping
operation, however, we do have to sample the Paulis such that their
first component is the identity. Doing so, we arrive at the tableau
shown in Figure~\ref{Fig:TableauB}(c). We then sweep the second random
pair of anticommuting Paulis to $X_2$ and $Z_2$, as shown in
Figure~\ref{Fig:TableauB}(d), and continue in a similar fashion until
we arrive at the final tableaus given by Figure~\ref{Fig:TableauA}(d)
and Figure~\ref{Fig:TableauA}(a).

In this procedure it can be seen that we operate on increasingly
smaller subtableaus consisting of two rows. The sweeping operations
have no effect on the entries outside the subtableau, since all
relevant entries are zero and therefore represent identity
Paulis. That means that we can start with an empty tableau with
exactly two rows, and perform all operations in-place on the same
tableau. 
Since every operation in the sweep routine corresponds to a gate in
the quantum circuit, we can output the circuit in a streaming fashion
while sweeping. In summary, the proposed algorithm performs the
following operations in successive iterations $\ell=1,\ldots,n$:

\medskip\noindent
{\bf{Step 1}} -- Consider the signed subtableau consisting of rows $2i-1$ and $2i$, and
  columns $x_j$ and $z_j$ for $j \geq \ell$. For the in-place
  version, simply restrict the columns of the two-row tableau.

\medskip\noindent
{\bf{Step 2}} -- Randomly sample two anticommuting Pauli operators on $n+1-\ell$
  qubits with random signs and assign them to the rows of the
  subtableau. This can be done using the algorithm described in
  Section~\ref{Sec:SamplingRows}.

\medskip\noindent
{\bf{Step 3}} -- Sweep the subtableau to basis Paulis $X_{1}$ and $Z_{1}$, which
  in the larger tableau corresponds to Paulis $X_{\ell}$ and
  $Z_{\ell}$. One possible
  algorithm for doing so is given in Section~\ref{Sec:Sweeping}.

  \medskip
  According to the discussion in Section~\ref{Sec:Motivation}, the
  above procedure outputs a quantum circuit corresponding to the
  adjoint of the randomly sampled Clifford operator
  $\mathcal{C}$. However, the adjoint of a Clifford operator is itself
  a Clifford operator, and given the one-to-one mapping between
  $\mathcal{C}$ and $\mathcal{C}^{\dag}$ it follows that the generated
  circuits are indeed sampled uniformly at random from the Clifford
  group. We prove correctness of the algorithm in
  Section~\ref{Sec:Correctness}.

\begin{figure}
\centering\setlength{\tabcolsep}{10pt}
\begin{tabular}{ccc}
\includegraphics[width=0.275\textwidth]{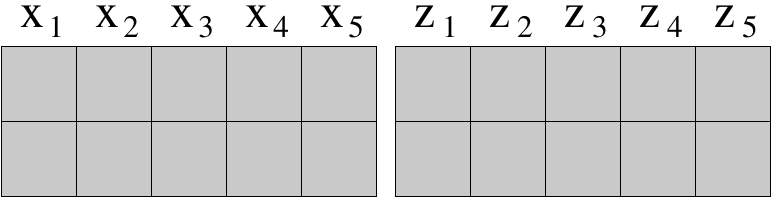}&
\includegraphics[width=0.275\textwidth]{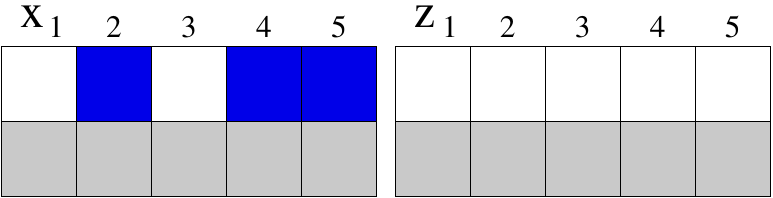}&
\includegraphics[width=0.275\textwidth]{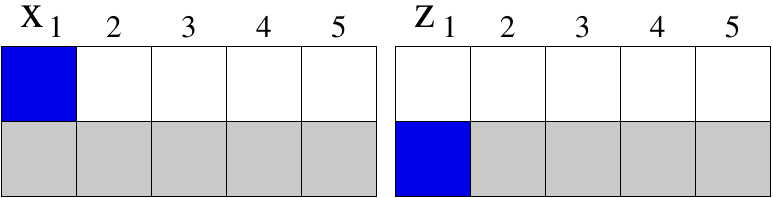}
\\
({\bf{a}}) & ({\bf{b}}) & ({\bf{c}})
\\[7pt]
\includegraphics[width=0.275\textwidth]{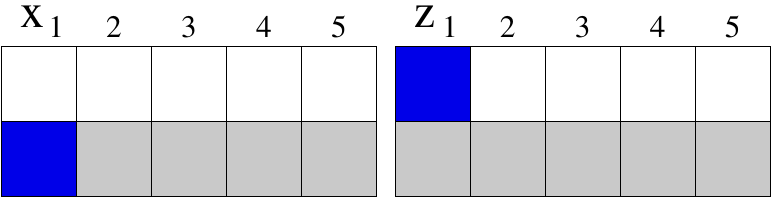}&
\includegraphics[width=0.275\textwidth]{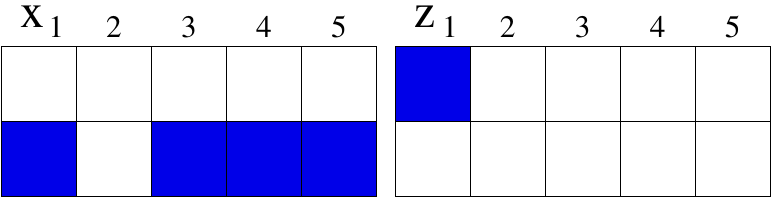}&
\includegraphics[width=0.275\textwidth]{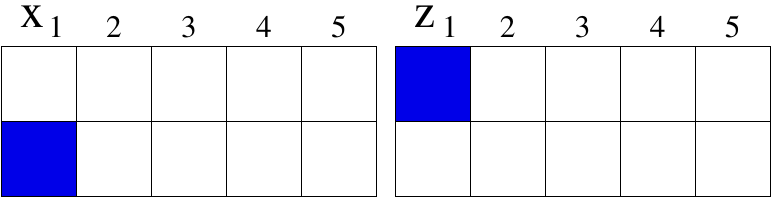}\\
({\bf{d}}) & ({\bf{e}}) & ({\bf{f}})
\end{tabular}
\caption{Illustration of the tableau at different stages of the sweeping process.}\label{Fig:TableauE}
\end{figure}

\subsection{Sampling rows}\label{Sec:SamplingRows}

As part of the proposed algorithm we need to randomly sample pairs of
anticommuting Pauli operators. For $k$ qubits we can sample individual
Pauli operators by randomly sampling $2k$ bits (a single row of the
tableau). There are a total of $4^k$ times $4^k$ possible Pauli pairs,
but not all pairs anti-commute. For the pair to be valid we require
that the first Pauli is not the identity, which gives $4^k-1$
options. For each such Pauli there are $4^k/2 = 2^{2k-1}$ Paulis that
anticommute. The probability of sampling a valid pair is therefore
\[
\frac{(4^k-1)2^{2k-1}}{4^{2k}} = (1 - 1/4^k)/2 \geq 3/8.
\]
If the pair is found to commute, we can simply discard them and repeat
the process. Depending on $k$, it takes between $2$ and $8/3$ trials
on average before we find an anticommuting pair. We have $k \leq n$,
and sampling the bits and determining commutativity therefore has an
expected time complexity of $\mathcal{O}(n)$. The algorithm has $n$
iterations, which thus gives an overall sample complexity of
$\mathcal{O}(n^2)$. This complexity matches the minimum, since we have
to sample at least $\log_2(\vert\mathcal{C}_n\vert)$ random bits,
where $\vert\mathcal{C}_n\vert$ is the cardinality of the Clifford
group given by equation~\eqref{Eq:CardinalityC}.

\subsection{Sweeping}\label{Sec:Sweeping}

For the sweeping step we are given a subtableau with anticommuting
$k$-Paulis $P(x^a,z^a)$ and $P(x^b,z^b)$, illustrated in
Figure~\ref{Fig:TableauE}(a) with separate $X$ and $Z$ blocks. The
goal of this step is to manipulate the tableau using the operations
from Figure~\ref{Fig:Tableau} to obtain a tableau that represents
Pauli operators $X_1$ and $Z_1$. The algorithm proceeds as follows:

\medskip\noindent {\bf{Step 1}} -- Clear the elements in the $Z$-block
of the first row. This is done by finding all indices $j$ for which
$z_j^a =1$ and applying H$(j)$ when $x_j^a = 0$, and S$(j)$
otherwise. This step generates a maximum of $k$ gates with circuit
depth one and has a time complexity of $\mathcal{O}(k)$. An example of
the result of this step is illustrated in
Figure~\ref{Fig:TableauE}(b).

\medskip\noindent
{\bf{Step 2}} -- Determine the (sorted) list of indices
$\mathcal{J} = \{j \mid x_j^a = 1\}$ where the $x^a$ coefficients are
nonzero. The set is guaranteed to be nonempty, otherwise the row would
correspond to the identity. Assuming one-based indexing, we can apply
CX$(\mathcal{J}_i,\mathcal{J}_{i+1})$ in parallel for all odd indices
$i < \vert\mathcal{J}\vert$ to clear up to half of the nonzero
coefficients in $x^a$. We then update $\mathcal{J}$ by retaining
only the values at the odd locations, namely
$\{\mathcal{J}_i \mid \mbox{odd}\ i\}$, and repeat the procedure until
$\mathcal{J}$ is a singleton. In the example in
Figure~\ref{Fig:TableauE}(b) we would start with
$\mathcal{J} = \{2,4,5\}$ and apply CX$(2,4)$ in the first stage. We
then update the index set to $\mathcal{J} = \{2,5\}$ and apply
CX$(2,5)$. One final update to the index set then gives
$\mathcal{J} = \{2\}$. This step has an $\mathcal{O}(k)$ time
complexity and generates a circuit with at most $k-1$ CX gates. The
circuit depth is bounded by $\lceil\log_2(k)\rceil$, which is the
maximum number of updates to $\mathcal{J}$.

\medskip\noindent
{\bf{Step 3}} -- When $\mathcal{J} \neq \{1\}$ we need to move the
remaining nonzero coefficient in $x^a$ to the first location. This is
done by swapping qubits $1$ and $\mathcal{J}_1$, which can be
implemented using three CX gates. The tableau at the end of this step
is as shown in Figure~\ref{Fig:TableauE}(c). For the second Pauli to
anticommute with $X_1$ it must have either $Y$ or $Z$ as the first
component. In either case that means that $z_1^b = 1$, as
indicated in the figure.

\medskip\noindent
{\bf{Step 4}} -- If the second Pauli is equal to $\pm Z_1$ we skip
this step. Otherwise we first apply the single-qubit gate H$(1)$ to
arrive a the tableau given by Figure~\ref{Fig:TableauE}(d). We then
repeat steps 1 and 2 with $x^a$ and $z^a$ replaced by $x^b$ and $z^b$:
we first use single-qubit operations to clear the elements in $z^b$,
as illustrated in Figure~\ref{Fig:TableauE}(e), and then zero out all but
one of the elements in $x^b$. The application of the Hadamard gate
ensures that the remaining element in the set $\mathcal{J}$ is 1, and
we are thus left with the tableau shown in
Figure~\ref{Fig:TableauE}(f). We again apply H$(1)$ to obtain Paulis
$X_1$ and $Z_1$.

\medskip\noindent {\bf{Step 5}} -- As a final step we clear any sign
bits by applying the appropriate Pauli operator on the first qubit. If
this operator anticommutes with the corresponding element in the Pauli
represented by the row it will cause the sign bit to flip. When
$s^a = 0$ and $s^b = 1$ we apply $X_1$ since it commutes with $X_1$
and anticommutes with $Z_1$. When $s^a = 1$ we apply $Y_1$ if $S^b=1$
and $Z_1$ otherwise. Note that, if we are interested only in the
operator, we could avoid maintaining signs throughout the sweeping and
simply sample the sign here to decide whether or not to apply the
above gates.

\medskip
The maximum number of single-qubit gates generate during the sweeping
process is $2k + 2$, while the maximum number of CX gates is
$2(k-1) + 3$. The circuit depth for fully connected qubits is at most
$8 + 2\lceil\log_2(k)\rceil$. Applying the sweeping step for $k$
ranging from $1$ to $n$ gives a maximum of $5n+2n^2$ gates
and a maximum circuit depth of
\[
\sum_{k=1}^n 8 + 2\lceil\log_2(k)\rceil \leq 10n + 2\log_2(n!) =
\mathcal{O}(n\log n).
\]

\begin{figure}
\centering
\includegraphics[width=0.55\textwidth]{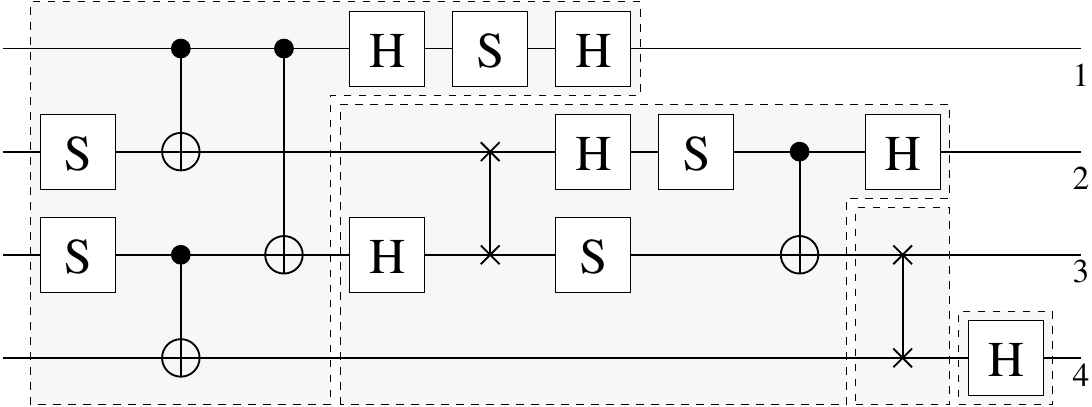}
\caption{Random four-qubit
  Clifford circuit generated by the proposed algorithm.}\label{Fig:Clifford}
\bigskip\medskip

\begin{tabular}{cccc}
\includegraphics[width=0.19\textwidth]{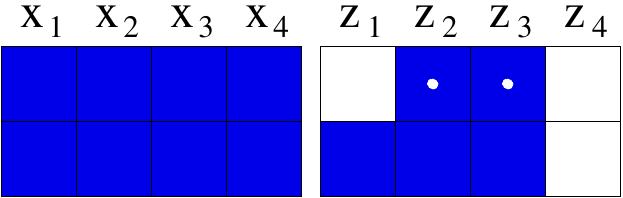}&
\includegraphics[width=0.19\textwidth]{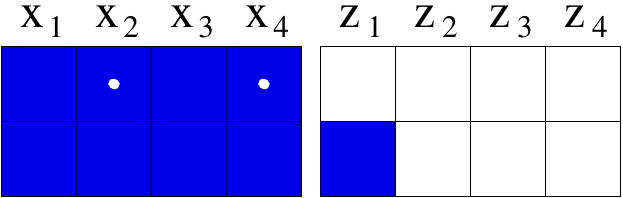}&
\includegraphics[width=0.19\textwidth]{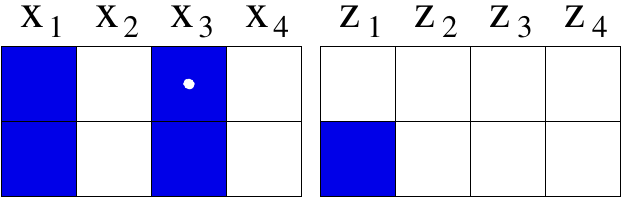}&
\includegraphics[width=0.19\textwidth]{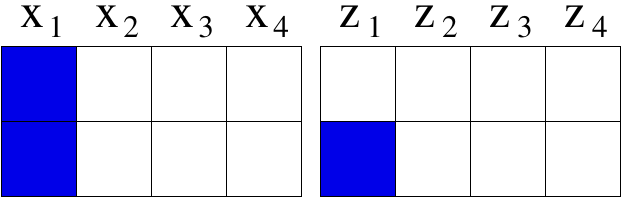}\\
({\bf{a}}) & ({\bf{b}}) & ({\bf{c}}) & ({\bf{d}})\\[3pt]
\includegraphics[width=0.19\textwidth]{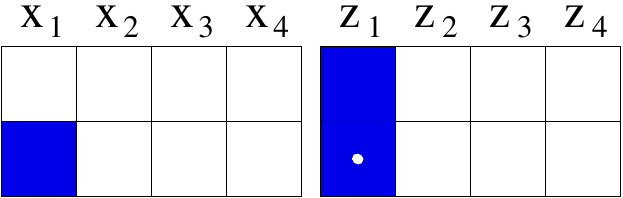}&
\includegraphics[width=0.19\textwidth]{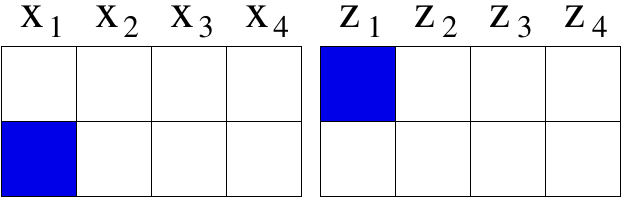}&
\includegraphics[width=0.19\textwidth]{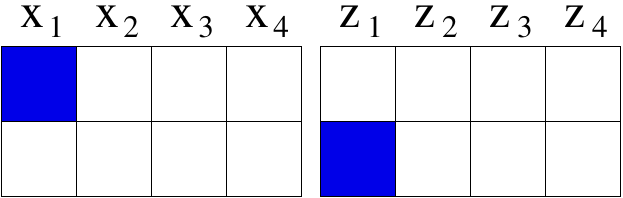}&
\includegraphics[width=0.19\textwidth]{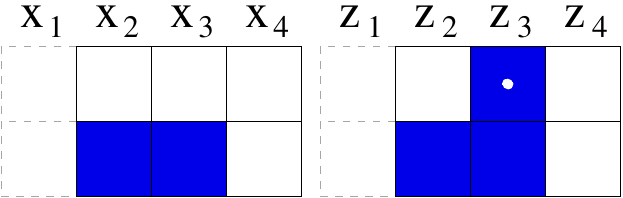}\\
({\bf{e}}) & ({\bf{f}}) & ({\bf{g}}) & ({\bf{h}})\\[3pt]
\includegraphics[width=0.19\textwidth]{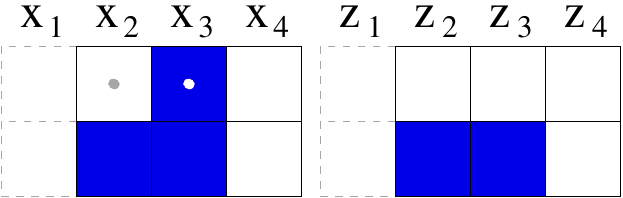}&
\includegraphics[width=0.19\textwidth]{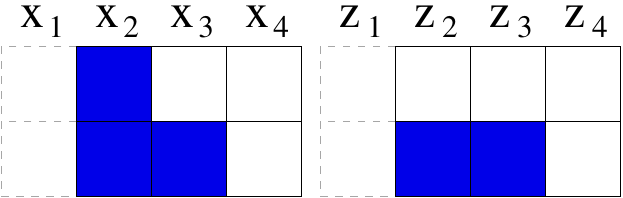}&
\includegraphics[width=0.19\textwidth]{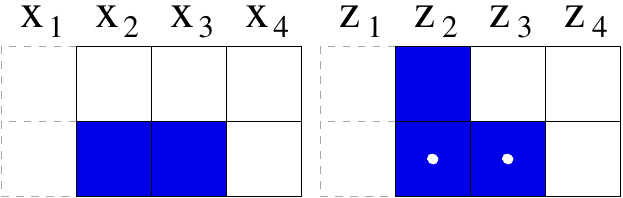}&
\includegraphics[width=0.19\textwidth]{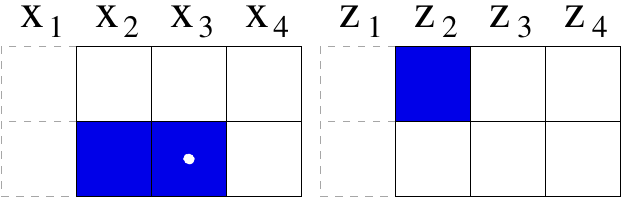}\\
({\bf{i}}) & ({\bf{j}}) & ({\bf{k}}) & ({\bf{l}})\\[3pt]
\includegraphics[width=0.19\textwidth]{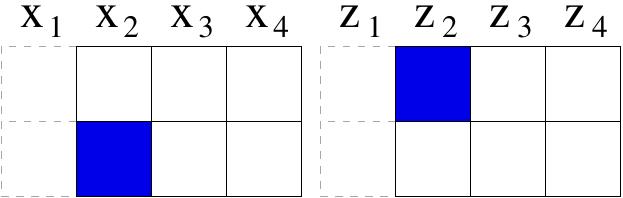}&
\includegraphics[width=0.19\textwidth]{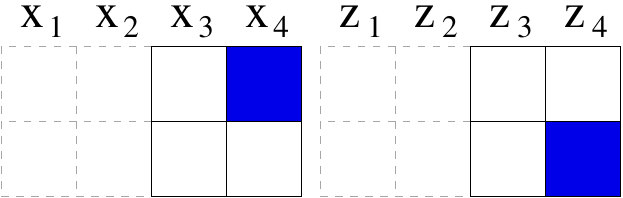}&
\includegraphics[width=0.19\textwidth]{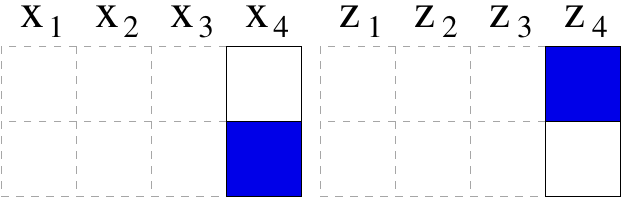}&\\
({\bf{m}}) & ({\bf{n}}) & ({\bf{o}})
\end{tabular}
\caption{Tableaus involved in generating the circuit in Figure~\ref{Fig:Clifford}; (a) the randomly sampled tableau at the first iteration of the algorithm. Step 1 of sweeping procedure applies phase gates to qubits 2 and 3 to clear entries $Z_2$ and $Z_3$ in the first row, as indicated by the white dots. The resulting tableau, shown in (b), has nonzero entries in $X_2$--$X_4$ of the first row, which we wish to clear in step 2. This is done pairwise, by applying CX gates on qubits (1,2) and (3,4), and then on (1,3), leading respectively to the tableaus shown in (c) and (d). Step 3 does not apply to (d), whereas application of the Hadamard gate in step 4 results in tableau (e). Sweeping steps 1 and 2 are then repeated for the second row. Step 1 clears $Z_1$ using a phase gate while, as seen in (f), no additional entries need to be cleared in step 2. As before, step 3 does not apply and step 4 applies another Hadamard to obtain the normalized tableau in (g). Although omitted in this example, Step 5 would then clear any sign bits by applying an appropriate Pauli gate. The gates corresponding to this first iteration are seen in the leftmost shaded region on qubits 1 through 4 in Figure~\ref{Fig:Clifford}. The second iteration of the sampling algorithm applies to qubits 2 through 4, and we start with the randomly sampled tableau shown in (h). Step 1, applied to the first row, applies a Hadamard gate to `exchange' the $Z_3$ and $X_3$ entries, resulting in tableau (i). Step 2 only finds a single nonzero entry in the X block and therefore does not apply any gates. Since the first entry in the tableau (corresponding to qubit 2) is zero, step 3 applies a swap operation to normalize the first row, as shown in  (j). Step 4 again applies a Hadamard gate to give tableau (k). We then repeat steps 1 and 2 on the second row. Step 1 clears the $Z_2$ and $Z_3$ entries, as indicated by the white dots, resulting in tableau (l). Step 2 then clears $X_3$ using a CX gate. Applying a Hadamard gate on the resulting tableau (m) again results in the normalized tableau, which we omit here. The randomly sampled tableau for iteration three is shown in (n). It can be seen that only the swap operation in step 3 applies. Finally, the random tableau for iteration four, shown in (o), is normalized by applying a Hadamard gate in sweeping step 1.}\label{Fig:CliffordSteps}
\end{figure}

Figure~\ref{Fig:Clifford} illustrates a random four-qubit Clifford circuit generated by the proposed algorithm. The corresponding tableaus, at various stages of the algorithm, are shown in Figure~\ref{Fig:CliffordSteps}, along with a detailed explanation in the caption.

\subsection{Correctness of the algorithm}\label{Sec:Correctness}

At iteration $k$ of the algorithm we randomly sample one of the
$2^{2k-1}(4^k-1)$ pairs of anticommuting Paulis. Adding signs
gives a total of $2^{2k+1}(4^k-1)$ different settings. Multiplying the
number of settings for iterations $k \in [1,n]$, we exactly obtain the
cardinality of the Clifford group given in
Equation~\eqref{Eq:CardinalityC}. Each of these settings has the same
probability of being sampled, and in order to prove that sampling is
uniform it therefore remains to show that each settings generates a
different Clifford operator.  Consider two settings whose sampled row
pairs first differ at iteration $k$, and denote the different pairs by
$(R_x,R_z)$ and $(\bar{R}_x,\bar{R}_z)$. Assume, without loss of
generality, that $R_x \neq \bar{R}_x$. Now, denote the Clifford
operator generated by the sweeping operations up to that point by
$\mathcal{C}$, which we define to be the identity operator if $k$ is
one. At iteration $k$ we sweep the tableau with operations that
correspond to operators $\mathcal{C}_k$ and
$\bar{\mathcal{C}}_k$. These operations are clearly different since
\[
\mathcal{C}_k^{\dag}X_k\mathcal{C}_k = R_x \neq
\bar{R}_x =\bar{\mathcal{C}}_k^{\dag}X_k\bar{\mathcal{C}}_k.
\]
The operations in subsequent iterations have no effect on $X_k$, and
the final Clifford operators therefore map $X_k$ to
$\mathcal{C}^{\dag}R_x\mathcal{C}$, and
$\mathcal{C}^{\dag}\bar{R}_x\mathcal{C}$, respectively. Conjugation
with Clifford operators never maps two Paulis to the same value, and we
therefore conclude that the sampled Clifford operators must
differ. This completes the proof.

\section{Discussion}\label{Sec:Discussion}

In this work we have proposed a simple algorithm for sampling
operators from the Clifford group uniformly at random. The algorithm
uses the tableau representation of Paulis and, similar to work by
Koenig and Smolin~\cite{KOE2014Sa}, takes advantage of the
hierarchical structure of the Clifford group. Unlike existing
algorithms, the algorithm can directly output the quantum circuit
corresponding to the random Clifford operator in a streaming fashion
with minimal overhead. The circuits generated for $n$-qubit Cliffords
contain at most $5n + 2n^2$ single- or two-qubit gates and have a
maximum depth of $\mathcal{O}(n \log n)$. The runtime of the algorithm
matches the $\mathcal{O}(n^2)$ complexity obtained by Bravyi and
Maslov~\cite{BRA2020Ma}. Each iteration of the algorithm consists of a
sampling step and a sweeping step. The sampling step randomly samples
a pair of anticommuting Paulis in tableau form. The sweeping step then
normalizes the tableau using operations that directly correspond to
elementary quantum gates. Each iteration of the algorithm therefore
results in a small quantum circuit.  By stitching together the
circuits generated by successive iterations we obtain a quantum
circuit that implements the randomly sampled Clifford operator.
Instead of stitching the circuits together at the end we can also emit
the gates as they are generated during the sweeping step and thus
output the circuit in a streaming manner. Interestingly, the
iterations in the algorithm are completely decoupled, which means that
the $n$ circuit segments could in principle be generated in parallel
in $\mathcal{O}(n)$ time.

Clifford operators are completely characterized by their mapping of the
basis Paulis $X_i$ and $Z_j$. If needed, we can obtain this mapping by
initializing the basis tableau shown in~\ref{Fig:TableauA}(a) and
applying the adjoint operations of each of the iterations, with the
iteration order reversed. The Paulis in the resulting tableau then
represent the exact mapping. This process requires operations on
increasingly large $k\times k$ subblocks of the tableau and results in
an $\mathcal{O}(n^3)$ algorithm, which matches the complexity
obtained by~\cite{KOE2014Sa}. The algorithm presented
in~\cite{BRA2020Ma} can find the same mapping with a complexity
matching that of matrix multiplication and is
therefore preferable, at least in theory.

\bibliography{bibliography}

\end{document}